\documentclass[10pt,preprintnumbers,superscriptaddress,amsmath,amssymb,aps, prb,showpacs]{revtex4-2}

\usepackage{stmaryrd}
\usepackage{amsmath}
\usepackage{amssymb}
\usepackage{float}
\usepackage{CJK}
\usepackage{array, color}
\usepackage{bm}
\usepackage{tabularx}
\usepackage{multirow}
\usepackage{threeparttable}
\usepackage{titletoc}
\usepackage{booktabs}
\usepackage{ulem}
\usepackage[colorlinks,linkcolor=blue,anchorcolor=blue,citecolor=blue,urlcolor=blue,driverfallback=dvipdfm]{hyperref}
\usepackage{lineno}
\usepackage{graphicx}
\usepackage{subfigure}
\usepackage{latexsym,bm}

\newcommand{\corre}{\mathrm{corre}}

\date{\today}
\begin{document}
\draft


\title{Thermal conductivity of MgO in giant planetary interior conditions \\ predicted by deep potential}


\author{Rong~Qiu}
\author{Qiyu~Zeng}
\author{Ke~Chen}
\author{Xiaoxiang~Yu}
  \email[Corresponding author: ]{xxyu@nudt.edu.cn}
\author{Jiayu~Dai}
  \email[Corresponding author: ]{jydai@nudt.edu.cn}
  \affiliation{Department of Physics, National University of Defense Technology, Changsha 410073, People's Republic of China}
  \affiliation{Hunan Key Laboratory of Extreme Matter and Applications, National University of Defense Technology, Changsha 410073, China}

\begin{abstract}
Thermal conductivity $\kappa$ of MgO plays a fundamental role in understanding the thermal evolution and mantle convection in the interior of terrestrial planets. However, previous theoretical calculations deviate from each other and the $\kappa$ of high-pressure B2 phase remains undetermined. Here, by combining molecular dynamics and deep potential trained with first-principles data, we systematically investigate the $\kappa$ of MgO from ambient state to the core-mantle boundary (CMB) of super-Earth with $5M_{\oplus}$. We point out the significance of 4-phonon scatterings and modify the conventional thermal conductivity model of MgO by considering the density-dependent proportion of 3-phonon and 4-phonon scatterings. The $\kappa$ profiles of MgO in Earth and super-Earth are further estimated. For super-Earth, we predict a significant reduction of $\kappa$ at the B1-B2 phase transition area near the CMB. This work provides new insights into thermal transport under extreme conditions and an improved thermal model for terrestrial planets.

\end{abstract}
\maketitle

\section{Introduction}

The conduction of heat into the overlying mantle at the core-mantle boundary (CMB) controls the thermal evolution of planet's core and history of the geomagnetic field \cite{ref:buffett2002}, which is of great scientific interest while highly challenging due to the extreme thermodynamic condition. Since magnesium oxide (MgO) is one of the most abundant materials in Earth's mantle \cite{ref:ringwood1991}, as well as major component in the rocky portions of other planets \cite{ref:valencia2006,ref:nettelmann2016,ref:wahl2016}, the thermal conductivity $\kappa$ of MgO serves as a fundamental role in understanding the heat balance, mantle convection, and thermal evolution of Earth and exoplanets \cite{ref:naliboff2006, ref:naliboff2007}.

On this issue, experimental efforts had been pursed to measure the $\kappa$ of MgO. However, available pressure (P) and temperature (T) conditions of experimental techniques are still far from the condition of CMB of Earth \cite{ref:goncharov2009,ref:dalton2013,ref:hasegawa2023}. Enormous theoretical researches have also been performed to predict the $\kappa$ of B1-phase MgO under extreme conditions. However, the reported values deviate significantly from each other.  From perspective of phonon gas theory, $\kappa$ can be theoretically estimated by solving the Boltzmann transport equation (BTE). Another approach applies molecular dynamics (MD) to calculate the $\kappa$ based on density functional theory (DFT) and classical interatiomic potential \cite{ref:stackhouse2010,ref:haigis2012}. Different functional forms in DFT and interatiomic potentials give different equation of states and $\kappa$ \cite{ref:tang2010,ref:kwon2020}. Moreover, the $\kappa$ estimated by considering different orders of phonon anharmonicity are also inconsistent \cite{ref:dekoker2010,ref:Ravichandran2019,ref:kwon2020}. Hence, there is still a lack of a sufficiently accurate result as a benchmark, as well as an accurate thermal conductivity model for practical geophysical applications. 

Besides, when considering the CMB condition for super-Earth with five times the mass of Earth ($5M_{\oplus}$), the pressure and temperature can reach as high as 650 GPa and 6000 K, respectively. Under such extreme condition, a transformation of MgO from NaCl-type (B1) to CsCl-type (B2) structure is expected to occur \cite{ref:soubiran2020}. To our knowledge, the $\kappa$ of B2-phase MgO remains unknown. 

Molecular dynamics inherently take into account full-order phonon anharmonicity, which avoid the considerable computation cost for including high-order phonon scatterings in BTE method. A key to accurately predict the $\kappa$ of MgO at extreme conditions is an interatomic potential with first-principles accuracy covering broad P-T ranges. Recent years, advances on deep learning techniques provides an efficient tool to solve the efficiency-accuracy dilemma in molecular modelling \cite{ref:zhang2018deep,ref:zeng2021ab,ref:zeng2022towards}. The deep neural network potential can reduce the computational complexity of \textit{ab initio} accurate inference of energies, interatomic forces and virial tensors to linear level. Especially, recent efforts have demonstrated the success of machine learning model towards large-scale simulations of thermal transport at extreme conditions with \textit{ab initio} quality \cite{ref:Deng2021,ref:Liu2021,ref:Wang2022,ref:Yang_2022,ref:Zhang2023thermal}.

In this work, we generate deep potential (DP) of MgO and validate its accuracy. Then we calculate the $\kappa$ of B1- and B2-phase MgO at extreme conditions up to 650 GPa and 6000 K and find a dramatic reduction of $\kappa$ across B1-B2 phase boundary. We demonstrate the importance of 4-phonon scatterings and point out the inapplicability of conventional thermal conductivity model. Through considering the density-dependent contributions of 3-phonon and 4-phonon scatterings, the fitting error of modified model decreases significantly. Finally, we report depth-dependent $\kappa$ profiles in the deep mantle of Earth and super-Earth.

\section{Method}

\subsection{Generation of deep potential}
  
 The DP models used in the DPMD simulations for MgO system are generated using DeePMD-kit packages \cite{dpkitv2}. To cover a wide range of thermodynamics conditions and minimize the computational consumption, a concurrent learning scheme, Deep Potential Generator (DP-GEN) \cite{ref:zhang2019active}, is adopted to sample the most compact and adequate data set. We choose B1-phase ($2\times 2\times 2$ supercells, containing 64 atoms) and B2-phase ($3\times 3\times 3$ supercells, containing 54 atoms) as the initial configurations and run DPMD under NPT ensemble covering temperature from 300 K to 7000 K and pressure from 0 GPa to 650 GPa. The training data sets contain 12472 configurations for B1 phase and 6738 configurations for B2 phase.

 The self-consistency calculations are performed using the VASP packages \cite{ref:KRESSE199615_vasp,ref:Kress_vasp}. The Perdew–Burke–Ernzerhof (PBE) parametrization of the generalized gradient approximation revised for solid (PBEsol) is implemented \cite{ref:perdew_1996,ref:perdew2008}, which can reproduce well both the thermal and the cold-curve equation of state of MgO, as shown in Fig. S1 in the Supporting Information (SI) or elsewhere \cite{ref:zhang2023}. The pseudopotential takes the projector augmented-wave (PAW) formalism \cite{ref:blochl1994projector,ref:holzwarth2001projector}. The kinetic energy cutoff is set to 900 eV and the sampling of the Brillouin zone is chosen as 0.5 ${\rm \mathring A^{-1}}$.

 For DP training, the embedding network is composed of three layers (25, 50, and 100 nodes) while the fitting network has three hidden layers with 240 nodes in each layer. The total number of training steps is set to $1\times 10^6$. The cutoff radius $r_c$ is set to $6.0\ {\rm \mathring A}$. The weight parameters in loss function for energies $p_e$, forces $p_f$, and virials $p_V$ are set to $(0.02, 1000, 0.02)$ at the beginning of training and gradually change to $(1.0, 1.0, 1.0)$. The accuracy of the DP model is tested in the whole sampling configuration. The fitting errors are shown in Fig. S2. For B1 phase, the RMSE of energy $\sigma_E$ is $6.29\ \rm{meV/atom}$, the relative RMSE (rRMSE) of forces, defined as $\sigma_f / \parallel f\parallel$ is $0.90\%$, and the RMSE of pressure is $0.20\ \rm{GPa}$. For B2 phase, the RMSE/rRMSE of energies, forces, and pressures is $3.43\ \rm{meV/atom}$, $0.99\%$, and $0.18\ \rm{GPa}$ respectively.

\subsection{Calculation of thermal conductivity}

Since MgO in giant planetary interior conditions remains a wide-band-gap insulator, heat conducts mainly via lattice vibration and the contribution from electrons can be neglected. Here, the heat current is calculated in DPMD simulations. The lattice thermal conductivity $\kappa$ is obtained from the integration of the heat current autocorrelation function (HCACF), known as Green–Kubo formula \cite{ref:mcquarrie1965statistical}, 
\begin{linenomath*}
\begin{equation}
\kappa_{\alpha\beta} = \frac{V}{k_\mathrm{B}T_i^2}\int_0^\infty \langle J_\alpha(0) J_\beta(t_{\corre}) \rangle dt_{\corre}
\end{equation}
\end{linenomath*}
where $\kappa_{\alpha \beta}$ is the $\alpha \beta^{th}$ component of the thermal conductivity tensor, $V$ is the volume of the material system, $k_B$ is the Boltzmann constant, $T_i$ is the lattice temperature, $t_{\corre}$ is the heat current autocorrelation time, and $J_\alpha$ is the $\alpha$-th component of the full heat current vector $\mathbf J$, which is typically computed as \cite{ref:plimpton1995fast},
\begin{linenomath*}
\begin{equation}
\mathbf J = \frac{1}{V} (\sum_i \mathbf v_i \epsilon_i + \sum_i \Xi_i \cdot \mathbf v_i)
\end{equation}
\end{linenomath*}
Here, $\mathbf v_i$, $\epsilon _i$ and $\Xi_i$ are the velocity, energy and stress tensor of atom $i$, respectively.
  
We performed all DPMD simulations with LAMMPS package \cite{ref:plimpton1995fast}. The size effect of $\kappa$ is shown in Fig. S4. For B1 phase under high pressure, a large supercell ($12\times 12\times 12$) containing 13842 atoms is used in DPMD simulations, which is almost unachievable for DFT calculations. The time step is set to 1 fs.  The Nos\'{e}-Hoover thermostat \cite{ref:nose1984unified,ref:hoover1985canonical} is employed in the NVT ensemble of 20 ps. Then, the ensemble is switched into NVE ensemble to calculate the HCACF in the next 200 ps with a correlation time of 20 ps. Every $\kappa$ value is the average of 20 individual cases with different initial velocity distribution.

\subsection{Lattice dynamics calculation}

We perform lattice dynamics calculations by using ALAMODE package \cite{ref:tadano2014}. The atomic forces are calculated in a $2\times 2 \times 2$ supercell for B1 phase and a $3\times 3 \times 3$ supercell for B2 phase. The finite displacement method with a displacement of 0.01 $\rm{\mathring A}$ is used to calculate the second-order interatomic force constants (IFCs) and phonon dispersion. To capture the LO-TO splitting, the Born effective charges are considered. 170 configurations with random displacement of 0.04 $\rm{\mathring A}$ are used to calculate the third-order IFCs and $Gr\ddot{u}neisen$ parameter ($\gamma$).

\section{Results}




\subsection{validity of neural network model}

\begin{figure}
\centering
  \includegraphics[width=1.0\textwidth]{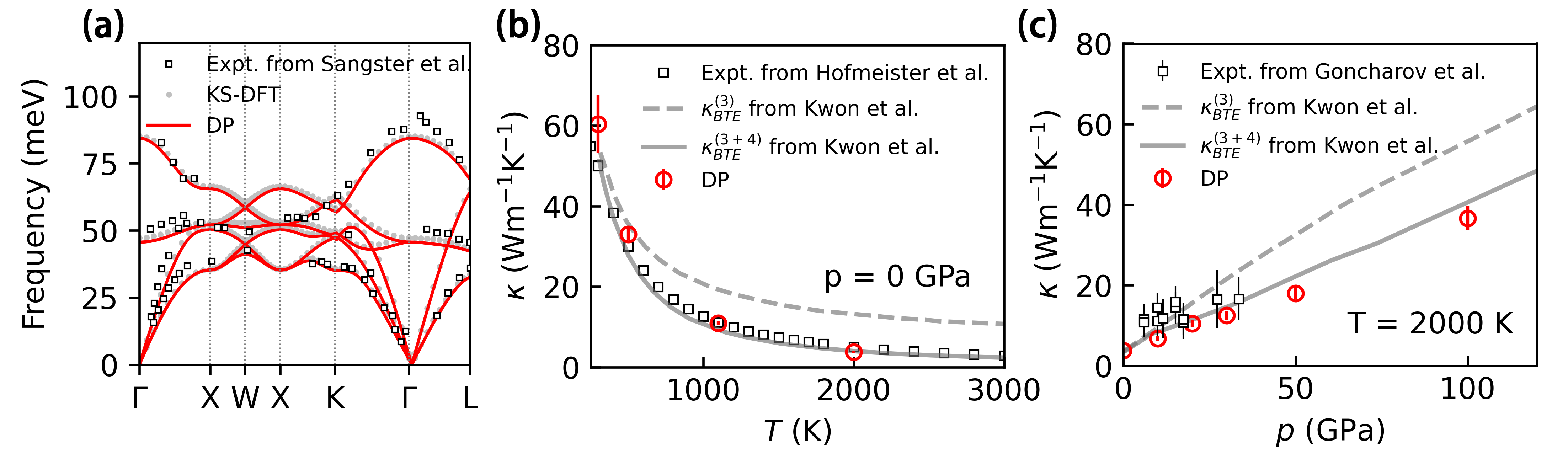}
  \caption{ (a) Phonon dispersion curves of B1-phase MgO. The red solid lines and gray dots represent the results calculated by DP and KS-DFT. The black squares represent experiment data from \cite{ref:Sangster1969}. The lattice parameter is set to $a_0$ = 4.238$\ \rm{\mathring A}$, corresponding to the condition of ambient pressure. Thermal conductivity $\kappa$ of B1-phase MgO (b) under P = 0 GPa and (c) along the isotherm of T = 2000 K. The DPMD results (red circles), BTE results including 3-phonon and 4-phonon scatterings \cite{ref:kwon2020} (gray dashed and solid lines), and experimental measurements\cite{ref:hofmeister2014,ref:goncharov2009} (black squares) are shown.}
  \label{fig:1}
\end{figure} 

Since the phonon dispersion serves as the key role in thermal transport, we first compare the phonon dispersion results obtained by different methods. As presented in the Fig.\ref{fig:1}(a), for B1-phase MgO under ambient pressure, the phonon dispersion curves calculated by DP agree well with KS-DFT calculations and experimental measurements \cite{ref:Sangster1969}. Phonon dispersion spectra of highly compressed B1-phase and B2-phase MgO is also reported to be consistent with individual $ab initio$ calculations (Fig. S3).

We also compare the $\kappa$ results in high P-T conditions obtained by different methods. Fig.\ref{fig:1} shows the $\kappa$ of B1-phase MgO under P = 0 GPa and along the isotherm of T = 2000 K. In previous BTE calculations, using DFT with PBEsol functional, the significance of 4-phonon scatterings in B1-phase MgO was pointed out and the inclusion of 4-phonon scatterings leads to the reduction of $\kappa$ \cite{ref:kwon2020,ref:Ravichandran2019}. As can be seen, the $\kappa$ predicted by DPMD is in good agreement with the BTE results including 4-phonon scatterings ($\kappa^{(3+4)}_{BTE}$) \cite{ref:kwon2020}. The good agreement can be attributed to the reason that full-order phonon anharmonicity is inherently included in MD simulations. Our DPMD results are also consistent with the experimental measurements\cite{ref:hofmeister2014,ref:goncharov2009}. Besides, due to the different settings of calculation details, there are several other calculations that deviates from our results (Fig. S5). \cite{ref:tang2010} and \cite{ref:dekoker2010} used local density approximation (LDA) functional and BTE including only 3-phonon scatterings. A recent work demonstrated that the LDA functional overestimates the density and predicts the EOS deviated from the experimental measurements \cite{ref:zhang2023}. \cite{ref:stackhouse2010} combined DFT method with non-equilibrium MD (NEMD) approach using LDA functional. However, the extrapolation method in NEMD simulations is only accurate when the system lengths are comparable or larger than the effective phonon mean free path \cite{ref:Sellan2010,ref:Dong2018}. Generally, the system size in first-principles MD simulations is much smaller than the phonon mean free path of B1-phase MgO. 
\cite{ref:haigis2012} calculated the $\kappa$ based on Green-Kubo approach with a empirical aspherical ion model potential, which was fitted from DFT datasets using LDA functional \cite{ref:Jahn2007}. 
The inaccuracy of LDA functional for MgO has been discussed above. In this work, the generated DP incorporates the DFT-PBEsol accuracy and full-order phonon anharmonicity into the large-scale calculation of $\kappa$. Therefore, our DPMD simulations can accurately predict the phonon properties and thermal transport of B1- and B2-phase MgO in broad P-T conditions.

\subsection{Thermal conductivity reduction at B1-B2 phase boundary}

\begin{figure}
\centering
  \includegraphics[width=1.0\textwidth]{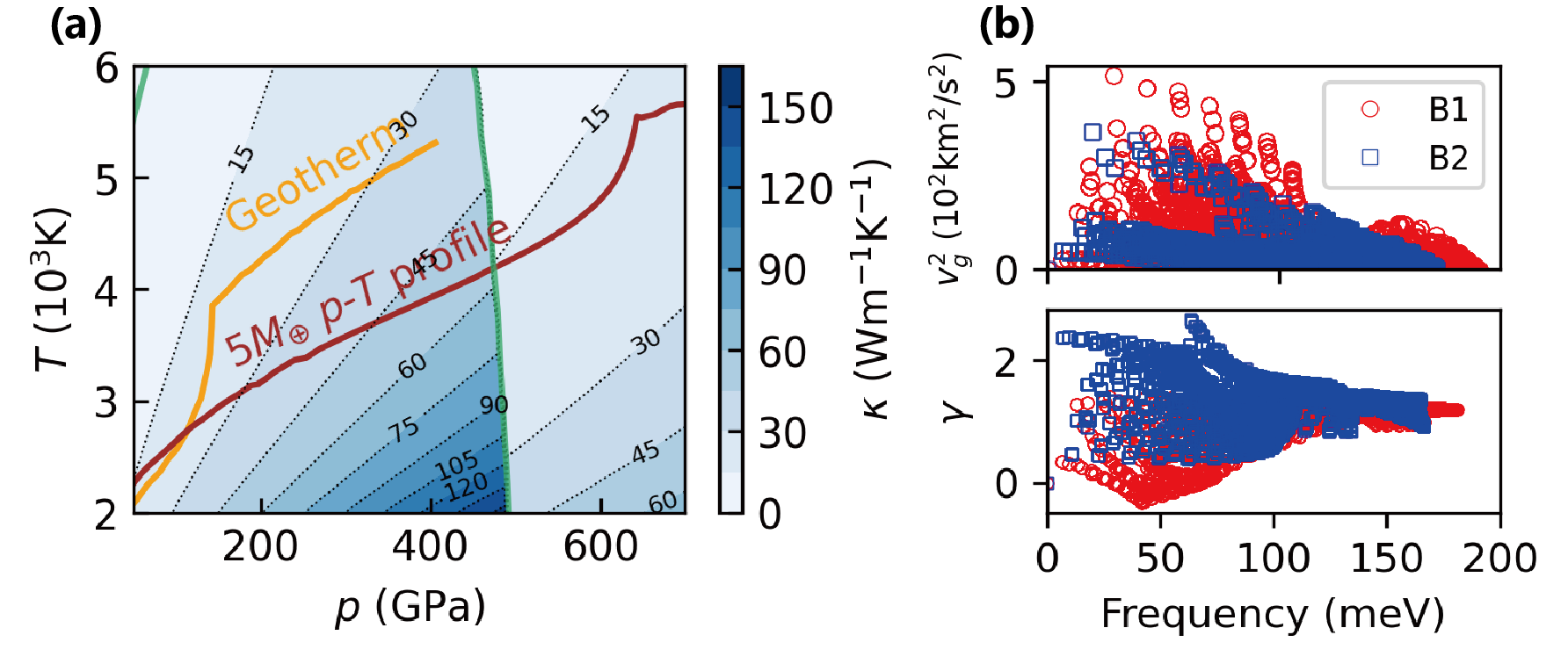}
  \caption{(a) The lattice thermal conductivity of MgO across the B1-B2 phase boundary. The B1-B2 phase boundary (green solid line) is obtained from Ref. \cite{ref:soubiran2020}. (b) Squared group velocity and (c) $Gr\ddot{u}neisen$ parameter $\gamma$ for B1 and B2 phase at pressure of 500 GPa.}
  \label{fig:2}
\end{figure}

We calculate the $\kappa$ of B2-phase MgO, as shown in Fig.~\ref{fig:2}(a). The B1-B2 phase boundary is referred from \cite{ref:soubiran2020}. It is clearly seen that the $\kappa$ of B2-phase MgO is much lower than that of B1-phase at the same P-T condition. As the sixfold-coordinated B1-phase structure changes into eightfold-coordinated B2-phase structure, the significant reduction of $\kappa$ is as high as 74.1\%, 69.9\%, and 64.7\% along the isotherm of T = 2000 K, 4000 K, and 6000 K, respectively (Fig. S6). 

The dramatic decrease can be understood by the phonon gas model, where the $\kappa$ is given as,
\begin{linenomath*}
\begin{equation}
\kappa = \sum_{i} c_i v_i^2 \tau_i
\end{equation}
\end{linenomath*}
$c_i$, $v_i$, and $\tau_i$ is the heat capacity, group velocity, and relaxation time of phonon mode $i$, respectively.
The reduction of $\kappa$ at the B1-B2 phase boundary is attributed to the decrease of group velocity and relaxation time. As shown in Fig.\ref{fig:2}(b), at 500 GPa, the squared group velocity $v^2$ of phonon modes in B1 phase is overall larger than that of B2 phase. Meanwhile, compared with B2 phase, the $Gr\ddot{u}neisen$ parameter $\gamma$ of phonon modes in B1 phase is overall smaller, indicating the stronger phonon anharmonicity and thus smaller relaxation time of phonon modes in B2 phase. Previous theoretical analysis also reported the sudden increase in phonon anharmonicity across B1-B2 structural transformation \cite{ref:slack1985}. Especially, the differences of low-frequency acoustic modes in B1- and B2-phase are more significant than those of high-frequency optical modes. The low-frequency acoustic modes was found to be dominant in heat conduction than the high-frequency optical modes and contribute about 87\% of the total $\kappa$ \cite{ref:tang2010}. 

\subsection{Modification of thermal conductivity model}

For the practical geophysical purpose, the variation of $\kappa$ with density and temperature are usually represented by \cite{ref:manthilake2011, ref:imada2014},
\begin{linenomath*}
\begin{equation}
\kappa = \kappa_0 (\frac{\rho}{\rho_0})^g (\frac{T_0}{T})^\beta
\end{equation}
\end{linenomath*}
where $g = b \ln(\frac{\rho}{\rho_0}) + c$, indicating that the $g$ is density-dependent value and cannot be set to a constant over the whole range of deep mantle \cite{ref:stackhouse2010}. Conventionally, $\beta$ is set to 1 \cite{ref:stackhouse2010,ref:dekoker2010,ref:tang2010}, because only 3-phonon scatterings are considered, which govern a temperature dependence of $\tau_3 \propto T^{-1}$ \cite{ref:roufosse1973}. However, as mentioned above, 4-phonon scatterings are non-negligible in MgO, which trigger a temperature dependence of $\tau_4 \propto T^{-2}$ \cite{ref:Feng2016}. Furthermore, as an ionic crystal with low mass ratio of 1.5, the contribution of acoustic-optic scattering would be considered important in MgO under high temperature \cite{ref:slack1985}. The combination of both 3-phonon and 4-phonon scatterings leads to $1 < \beta < 2$. Therefore, for the $\kappa$ of MgO, $\beta$ can not be set to be 1 as used in previous works. Moreover, previous work reported the varying contributions of 3-phonon and 4-phonon scatterings in MgO at different pressures \cite{ref:Ravichandran2019}, suggesting that $\beta$ varies with density. Hence $\beta$ can not be set to a constant especially in the deep mantle where density of minerals grows with increasing pressure. 

\begin{table*}[htbp]
\caption{Fitting parameters of $\kappa(\rho, T)$. The fitting error is quantified by relative RMSE. Bold numbers correspond to lower values. }
\centering
\begin{tabular}{cccccccc}
 \hline
 Fitting setting & Phase & $b$ & $c$ & $\beta$ & $e$ & $f$ & rRMSE \\
 \hline
 \multirow{2}{*}{$\beta$=1} & B1 & -1.68 & 5.20 & 1.00 & - & - & 23.10\% \\
 & B2 & 18.26 & 2.18 & 1.00 & - & - &  7.43\% \\
 \hline
 \multirow{2}{*}{$\beta\neq$1} & B1 & -3.58 & 7.52 & 1.33 & - & - & 10.89\% \\
 & B2 & 13.73 & 2.85 & 1.13 & - & - &  2.92\% \\
 \hline
 \multirow{2}{*}{$\beta(\rho)$} & B1 & -2.33 & 6.96 & - & 0.48 & 1.11 & \textbf{5.33\%} \\
 & B2 & 13.14 & 2.93 & - & 0.46 & 1.10 & \textbf{2.66\%} \\
 \hline
 \label{tab:1}
\end{tabular}
\vspace*{-4pt}
\end{table*}

Analogous to $g$, we propose modifying the model by setting $\beta = e \ln(\frac{\rho}{\rho_0}) + f$ to consider the density dependence of $\beta$. The $\kappa_0$, $\rho_0$, $T_0$ are chosen as $55.20\ \rm{Wm^{-1}K^{-1}}$, $3.54\ \rm{g/cm^{3}}$, and 300 K for B1 phase (p = 0 GPa, T = 300 K), and $30.88\ \rm{Wm^{-1}K^{-1}}$, $7.09\ \rm{g/cm^{3}}$, and 2000 K for B2 phase (p = 400 GPa, T = 2000 K). Three fitting settings of $\beta$ are compared. $\beta$ = 1, $\beta \neq$ 1, and $\beta(\rho)$ represent constant $\beta$ equals to 1, constant $\beta$ not equals to 1, and $\beta$ that varies with $\rho$, respectively. The fitting parameters are listed in the Table \ref{tab:1}. When $\beta$ = 1, the large fitting error proves that only considering 3-phonon scatterings is inaccurate. When $\beta \neq$ 1, the inclusion of 4-phonon scatterings reduces the fitting error. The average $\beta$ under different $\rho$ conditions are obtained as 1.33 and 1.13 for B1 and B2 phase respectively, pointing out the larger proportion of 4-phonon scatterings in B1 phase. When $\beta$ varies with $\rho$, the fitting errors for B1 and B2 are as small as 5.33\% and 2.66\% respectively. $\beta(\rho)$ shows the best fitting compared with $\beta$ = 1 and $\beta \neq$ 1 (Fig. S7). The significantly decreased fitting errors indicate that the modified model can well capture the density-dependent contributions of 3-phonon and 4-phonon scatterings.

\begin{figure}
\centering
  \includegraphics[width=0.5\textwidth]{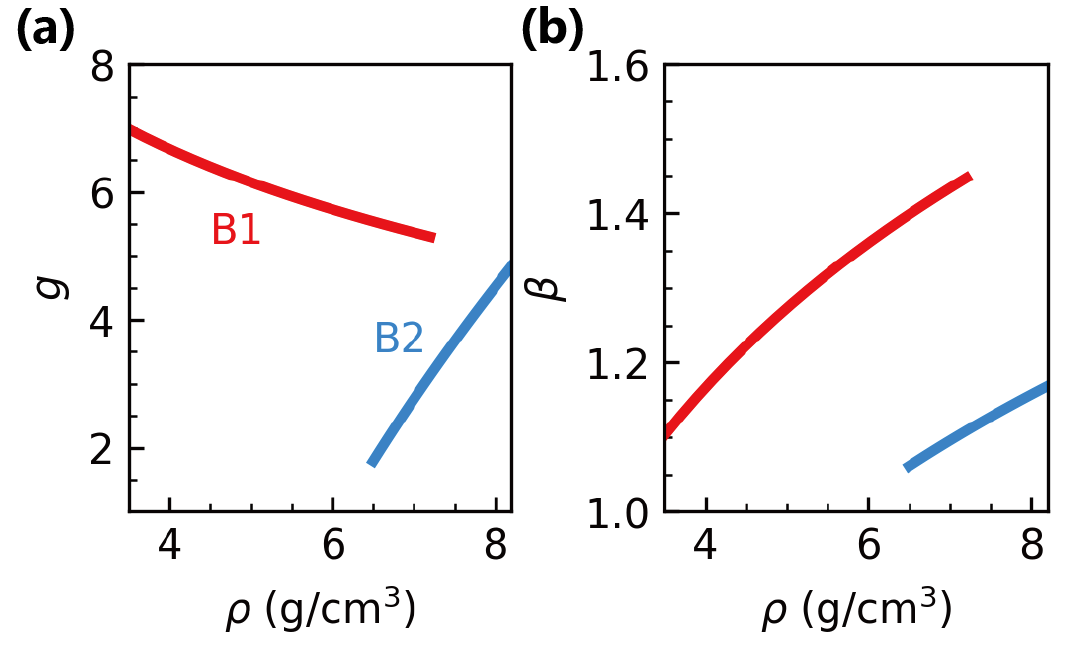}
  \caption{(a) $g$ and (b) $\beta$ in modified thermal conductivity model as functions of density.}
  \label{fig:3}
\end{figure}

To gain further insight into the density dependence of $\kappa$, the fitting parameter $g$ and $\beta$ as functions of density are showed in Fig.~\ref{fig:3}. The positive value of $g$ means a positive dependence of $\kappa$ versus $\rho$. The $g$ for B1 phase is larger than that for B2 phase, suggesting a sharper increase tendency of $\kappa$ for B1 phase. The values of $\beta$ for both B1 and B2 phase are larger than 1, which is due to the non-negligible contribution of 4-phonon scatterings as explained above. The positive dependence of $\beta$ on density demonstrates the increasing proportion of 4-phonon scatterings with increasing density. Compared with B2 phase, the larger $\beta$ for B1 phase represents the larger proportion of 4-phonon scatterings in B1 phase.

\subsection{Depth-dependent $\kappa$ along the planetary thermodynamic profile}

\begin{figure}
\centering
  \includegraphics[width=1.0\textwidth]{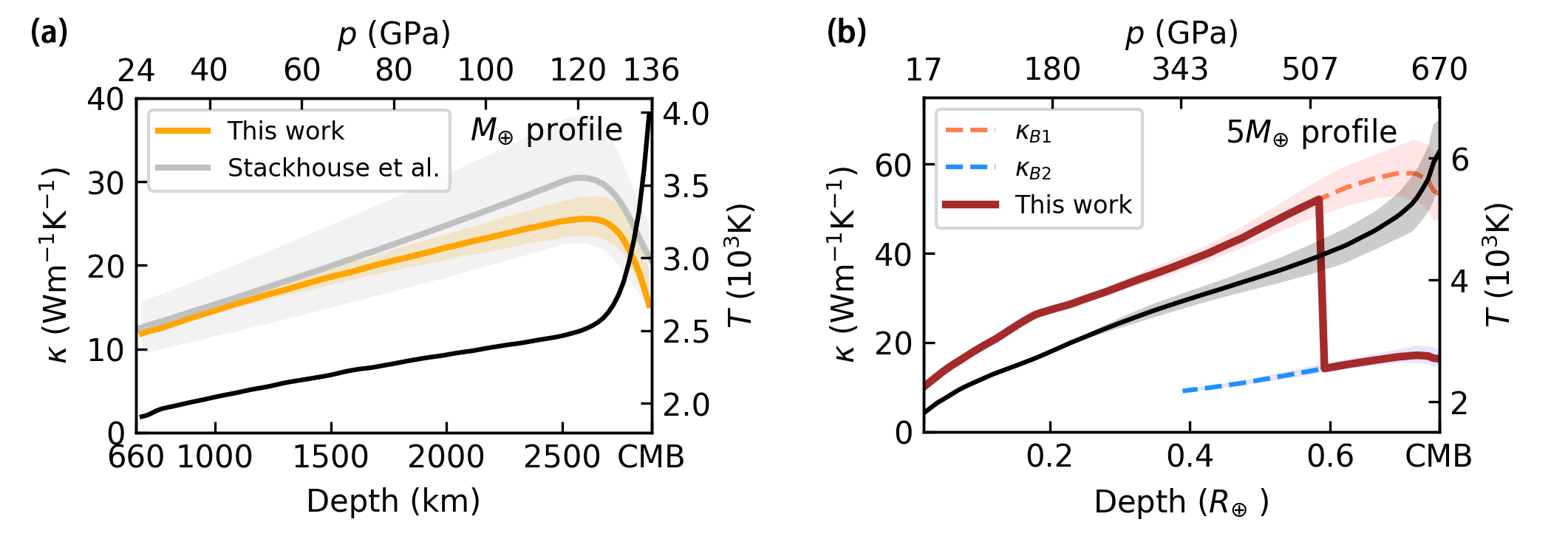}
  \caption{(a) $\kappa$ of B1-phase MgO along the geotherm. Previous estimation of $\kappa$ based on NEMD approach \cite{ref:stackhouse2010} are presented for comparison. (b) Depth-dependent $\kappa$ of MgO along the thermodynamic profile of $5M_{\oplus}$ exoplanets with Earth-like bulk composition \cite{ref:wagner2011}.}
  \label{fig:4}
\end{figure}

Having a robust model of the variation of $\kappa$ with pressure/density and temperature, here we provide an estimation of depth-dependent $\kappa$ values in the interior of Earth and Earth-like exoplanets. As shown in Fig.~\ref{fig:4}. The $\kappa$ along a lower mantle geotherm is estimated based on the DPMD-predicted $\kappa(p,T)$, with the depth-dependent pressure taken from the Preliminary Reference Earth Model (PREM) \cite{ref:dziewonski1981} and the temperature profile adopted from \cite{ref:stixrude2007}. Fig.\ref{fig:4}(a) shows the $\kappa$ along the geotherm close to the CMB of Earth. With increasing depth, the pressure effect overwhelms the concomitant temperature rise. $\kappa$ increases from $11.84\ \rm{Wm^{-1}K^{-1}}$ at the top of the lower mantle (660 km) to $25.56\ \rm{Wm^{-1}K^{-1}}$ at the top of the thermal boundary layer (2685 km) above the core-mantle boundary (2885 km). In comparison, \cite{ref:stackhouse2010} reported a higher value of $30.47\ \rm{Wm^{-1}K^{-1}}$ at the depth of 2685 km. The discrepancy can be attributed to the choice of LDA functional and the $1/T$ dependency of $\kappa$ in Stackhouse et al.'s work, as discussed in previous section 3.1. At the CMB condition (p = 136 GPa, T = 4100 K) at the top of the outer core, our results give a $\kappa$ of $15.81_{-0.98}^{+5.31}\ \rm{Wm^{-1}K^{-1}}$ when considering the temperature variation of $4100^{-800}_{+200}\ \rm{K}$. 

Such B1-B2 phase transition is expected to widely exist in the deep mantle of massive terrestrial planets due to the extreme thermodynamic condition.
Here we gives the depth-dependent $\kappa$ along the thermodynamic profile of $5M_{\oplus}$ exoplanets with Earth-like bulk composition (planetary radius is $\sim 1.55R_{\oplus}$) \cite{ref:wagner2011}. The pressure and temperature at the CMB (depth $\sim 0.76 R_{\oplus}$) is estimated to be $645\sim710\ \rm{GPa}$ and $5510\sim6470\ \rm{K}$, exceeding the B1-B2 phase transition boundaries.
As shown in Fig.\ref{fig:4}(b), with increasing depth, a maximum $\kappa$ of $51.74\ \rm{Wm^{-1}K^{-1}}$ at $0.58 R_{\oplus}$ (p = 465 GPa, T = 4388 K) and then a dramatic decrease of $\sim73.6\%$ into $14.18\ \rm{Wm^{-1}K^{-1}}$ across the B1-B2 phase boundary are observed. Further, an increasingly super-adiabatic temperature rise towards the CMB leads to that the $\kappa$ of B2 phase increases slowly due to the competition between positive effect of pressure and negative effect of temperature. At the top of metallic core, a maximum value of $17.13\ \rm{Wm^{-1}K^{-1}}$ is achieved.
We note that, the sharp change of $\kappa$ from high-$\kappa$ B1 phase to low-$\kappa$ B2 phase would have significant influence on the heat transfer of the super-Earth's mantle.

\section{Conclusion} 

In summary, by utilizing the DPMD and Green-Kubo approach, we systematically investigate the $\kappa$ of MgO under exoplanet conditions within the accuracy of PBEsol functional.
We show that the DPMD approach inherently captures the full-order phonon anharmornicity and predicted $\kappa$ in good agreement with previous BTE results including 3-phonon and 4-phonon scatterings and with experiment measurements. The $\kappa$ of both B1 and B2 phase up to the CMB condition of super-Earth with $5M_{\oplus}$ are calculated. It is found that the $\kappa$ exhibits an obvious increasing trend with increasing pressure. A significant reduction of $\kappa$ is observed across the B1-B2 phase transition. Lattice dynamics analysis show the reduction originates from the decreased group velocity and increased $Gr\ddot{u}neisen$ parameter. We modify the thermal conductivity model by change the $\beta$ from 1 to a density-dependent variable. The significantly reduced fitting error demonstrates the necessity of including 4-phonon scatterings and density-dependent ratio of 3-phonon and 4-phonon scatterings. Based on modified model, we further estimate the $\kappa$ of MgO along the thermodynamic profile of Earth and super-Earth. Especially for super Earth, a decrease of $\kappa$ from $51.74\ \rm{Wm^{-1}K^{-1}}$ to $14.18\ \rm{Wm^{-1}K^{-1}}$ across B1-B2 phase transition at depth of $0.58R_{\oplus}$ is predicted. Our calculations about $\kappa$ of MgO and modified model would serve an important role for understanding the heat transfer in the deep mantle of massive terrestrial planets, and the cores of icy and gas giants.

\section*{Acknowledgment}
    the Science and Technology Innovation Program of Hunan Province under Grant No. 2021RC4026, the National Natural Science Foundation of China under Grant No. 12304307.

\bibliography{reference}

\begin{thebibliography}{52}%
\makeatletter
\providecommand \@ifxundefined [1]{%
 \@ifx{#1\undefined}
}%
\providecommand \@ifnum [1]{%
 \ifnum #1\expandafter \@firstoftwo
 \else \expandafter \@secondoftwo
 \fi
}%
\providecommand \@ifx [1]{%
 \ifx #1\expandafter \@firstoftwo
 \else \expandafter \@secondoftwo
 \fi
}%
\providecommand \natexlab [1]{#1}%
\providecommand \enquote  [1]{``#1''}%
\providecommand \bibnamefont  [1]{#1}%
\providecommand \bibfnamefont [1]{#1}%
\providecommand \citenamefont [1]{#1}%
\providecommand \href@noop [0]{\@secondoftwo}%
\providecommand \href [0]{\begingroup \@sanitize@url \@href}%
\providecommand \@href[1]{\@@startlink{#1}\@@href}%
\providecommand \@@href[1]{\endgroup#1\@@endlink}%
\providecommand \@sanitize@url [0]{\catcode `\\12\catcode `\$12\catcode
  `\&12\catcode `\#12\catcode `\^12\catcode `\_12\catcode `\%12\relax}%
\providecommand \@@startlink[1]{}%
\providecommand \@@endlink[0]{}%
\providecommand \url  [0]{\begingroup\@sanitize@url \@url }%
\providecommand \@url [1]{\endgroup\@href {#1}{\urlprefix }}%
\providecommand \urlprefix  [0]{URL }%
\providecommand \Eprint [0]{\href }%
\providecommand \doibase [0]{https://doi.org/}%
\providecommand \selectlanguage [0]{\@gobble}%
\providecommand \bibinfo  [0]{\@secondoftwo}%
\providecommand \bibfield  [0]{\@secondoftwo}%
\providecommand \translation [1]{[#1]}%
\providecommand \BibitemOpen [0]{}%
\providecommand \bibitemStop [0]{}%
\providecommand \bibitemNoStop [0]{.\EOS\space}%
\providecommand \EOS [0]{\spacefactor3000\relax}%
\providecommand \BibitemShut  [1]{\csname bibitem#1\endcsname}%
\let\auto@bib@innerbib\@empty
\bibitem [{\citenamefont {Buffett}(2002)}]{ref:buffett2002}%
  \BibitemOpen
  \bibfield  {author} {\bibinfo {author} {\bibfnamefont {B.~A.}\ \bibnamefont
  {Buffett}},\ }\bibfield  {title} {\bibinfo {title} {Estimates of heat flow in
  the deep mantle based on the power requirements for the geodynamo},\ }\href
  {https://doi.org/10.1029/2001gl014649} {\bibfield  {journal} {\bibinfo
  {journal} {Geophysical Research Letters}\ }\textbf {\bibinfo {volume} {29}},\
  \bibinfo {pages} {7174} (\bibinfo {year} {2002})}\BibitemShut {NoStop}%
\bibitem [{\citenamefont {Ringwood}(1991)}]{ref:ringwood1991}%
  \BibitemOpen
  \bibfield  {author} {\bibinfo {author} {\bibfnamefont {A.~E.}\ \bibnamefont
  {Ringwood}},\ }\bibfield  {title} {\bibinfo {title} {Phase transformations
  and their bearing on the constitution and dynamics of the mantle},\ }\href
  {https://doi.org/10.1016/0016-7037(91)90090-R} {\bibfield  {journal}
  {\bibinfo  {journal} {Geochim. Cosmochim. Acta}\ }\textbf {\bibinfo {volume}
  {55}},\ \bibinfo {pages} {2083} (\bibinfo {year} {1991})}\BibitemShut
  {NoStop}%
\bibitem [{\citenamefont {Valencia}\ \emph {et~al.}(2006)\citenamefont
  {Valencia}, \citenamefont {O'Connell},\ and\ \citenamefont
  {Sasselov}}]{ref:valencia2006}%
  \BibitemOpen
  \bibfield  {author} {\bibinfo {author} {\bibfnamefont {D.}~\bibnamefont
  {Valencia}}, \bibinfo {author} {\bibfnamefont {R.~J.}\ \bibnamefont
  {O'Connell}},\ and\ \bibinfo {author} {\bibfnamefont {D.}~\bibnamefont
  {Sasselov}},\ }\bibfield  {title} {\bibinfo {title} {Internal structure of
  massive terrestrial planets},\ }\href
  {https://doi.org/10.1016/j.icarus.2005.11.021} {\bibfield  {journal}
  {\bibinfo  {journal} {Icarus}\ }\textbf {\bibinfo {volume} {181}},\ \bibinfo
  {pages} {545} (\bibinfo {year} {2006})}\BibitemShut {NoStop}%
\bibitem [{\citenamefont {Nettelmann}\ \emph {et~al.}(2016)\citenamefont
  {Nettelmann}, \citenamefont {Wang}, \citenamefont {Fortney}, \citenamefont
  {Hamel}, \citenamefont {Yellamilli}, \citenamefont {Bethkenhagen},\ and\
  \citenamefont {Redmer}}]{ref:nettelmann2016}%
  \BibitemOpen
  \bibfield  {author} {\bibinfo {author} {\bibfnamefont {N.}~\bibnamefont
  {Nettelmann}}, \bibinfo {author} {\bibfnamefont {K.}~\bibnamefont {Wang}},
  \bibinfo {author} {\bibfnamefont {J.~J.}\ \bibnamefont {Fortney}}, \bibinfo
  {author} {\bibfnamefont {S.}~\bibnamefont {Hamel}}, \bibinfo {author}
  {\bibfnamefont {S.}~\bibnamefont {Yellamilli}}, \bibinfo {author}
  {\bibfnamefont {M.}~\bibnamefont {Bethkenhagen}},\ and\ \bibinfo {author}
  {\bibfnamefont {R.}~\bibnamefont {Redmer}},\ }\bibfield  {title} {\bibinfo
  {title} {Uranus evolution models with simple thermal boundary layers},\
  }\href {https://doi.org/10.1016/j.icarus.2016.04.008} {\bibfield  {journal}
  {\bibinfo  {journal} {Icarus}\ }\textbf {\bibinfo {volume} {275}},\ \bibinfo
  {pages} {107} (\bibinfo {year} {2016})}\BibitemShut {NoStop}%
\bibitem [{\citenamefont {Wahl}\ \emph {et~al.}(2017)\citenamefont {Wahl},
  \citenamefont {Hubbard}, \citenamefont {Kaspi}, \citenamefont {Connerney},
  \citenamefont {Helled}, \citenamefont {Reese}, \citenamefont {Militzer},
  \citenamefont {Guillot}, \citenamefont {Miguel}, \citenamefont {Levin},
  \citenamefont {Galanti},\ and\ \citenamefont {Bolton}}]{ref:wahl2016}%
  \BibitemOpen
  \bibfield  {author} {\bibinfo {author} {\bibfnamefont {S.~M.}\ \bibnamefont
  {Wahl}}, \bibinfo {author} {\bibfnamefont {W.~B.}\ \bibnamefont {Hubbard}},
  \bibinfo {author} {\bibfnamefont {Y.}~\bibnamefont {Kaspi}}, \bibinfo
  {author} {\bibfnamefont {J.~E.}\ \bibnamefont {Connerney}}, \bibinfo {author}
  {\bibfnamefont {R.}~\bibnamefont {Helled}}, \bibinfo {author} {\bibfnamefont
  {D.}~\bibnamefont {Reese}}, \bibinfo {author} {\bibfnamefont
  {B.}~\bibnamefont {Militzer}}, \bibinfo {author} {\bibfnamefont
  {T.}~\bibnamefont {Guillot}}, \bibinfo {author} {\bibfnamefont
  {Y.}~\bibnamefont {Miguel}}, \bibinfo {author} {\bibfnamefont
  {S.}~\bibnamefont {Levin}}, \bibinfo {author} {\bibfnamefont
  {E.}~\bibnamefont {Galanti}},\ and\ \bibinfo {author} {\bibfnamefont {S.~J.}\
  \bibnamefont {Bolton}},\ }\bibfield  {title} {\bibinfo {title} {Comparing
  jupiter interior structure models to juno gravity measurements and the role
  of a dilute core},\ }\href {https://doi.org/10.1002/2017GL073160} {\bibfield
  {journal} {\bibinfo  {journal} {Geophysical Research Letters}\ }\textbf
  {\bibinfo {volume} {44}},\ \bibinfo {pages} {4649} (\bibinfo {year}
  {2017})}\BibitemShut {NoStop}%
\bibitem [{\citenamefont {Naliboff}\ and\ \citenamefont
  {Kellogg}(2006)}]{ref:naliboff2006}%
  \BibitemOpen
  \bibfield  {author} {\bibinfo {author} {\bibfnamefont {J.~B.}\ \bibnamefont
  {Naliboff}}\ and\ \bibinfo {author} {\bibfnamefont {L.~H.}\ \bibnamefont
  {Kellogg}},\ }\bibfield  {title} {\bibinfo {title} {Dynamic effects of a
  step-wise increase in thermal conductivity and viscosity in the lowermost
  mantle},\ }\href {https://doi.org/10.1029/2006gl025717} {\bibfield  {journal}
  {\bibinfo  {journal} {Geophysical Research Letters}\ }\textbf {\bibinfo
  {volume} {33}} (\bibinfo {year} {2006})}\BibitemShut {NoStop}%
\bibitem [{\citenamefont {Naliboff}\ and\ \citenamefont
  {Kellogg}(2007)}]{ref:naliboff2007}%
  \BibitemOpen
  \bibfield  {author} {\bibinfo {author} {\bibfnamefont {J.~B.}\ \bibnamefont
  {Naliboff}}\ and\ \bibinfo {author} {\bibfnamefont {L.~H.}\ \bibnamefont
  {Kellogg}},\ }\bibfield  {title} {\bibinfo {title} {Can large increases in
  viscosity and thermal conductivity preserve large-scale heterogeneity in the
  mantle?},\ }\href {https://doi.org/10.1016/j.pepi.2007.01.009} {\bibfield
  {journal} {\bibinfo  {journal} {Physics of the Earth and Planetary
  Interiors}\ }\textbf {\bibinfo {volume} {161}},\ \bibinfo {pages} {86}
  (\bibinfo {year} {2007})}\BibitemShut {NoStop}%
\bibitem [{\citenamefont {Goncharov}\ \emph {et~al.}(2009)\citenamefont
  {Goncharov}, \citenamefont {Beck}, \citenamefont {Struzhkin}, \citenamefont
  {Haugen},\ and\ \citenamefont {Jacobsen}}]{ref:goncharov2009}%
  \BibitemOpen
  \bibfield  {author} {\bibinfo {author} {\bibfnamefont {A.~F.}\ \bibnamefont
  {Goncharov}}, \bibinfo {author} {\bibfnamefont {P.}~\bibnamefont {Beck}},
  \bibinfo {author} {\bibfnamefont {V.~V.}\ \bibnamefont {Struzhkin}}, \bibinfo
  {author} {\bibfnamefont {B.~D.}\ \bibnamefont {Haugen}},\ and\ \bibinfo
  {author} {\bibfnamefont {S.~D.}\ \bibnamefont {Jacobsen}},\ }\bibfield
  {title} {\bibinfo {title} {Thermal conductivity of lower-mantle minerals},\
  }\href {https://doi.org/10.1016/j.pepi.2008.07.033} {\bibfield  {journal}
  {\bibinfo  {journal} {Physics of the Earth and Planetary Interiors}\ }\textbf
  {\bibinfo {volume} {174}},\ \bibinfo {pages} {24} (\bibinfo {year}
  {2009})}\BibitemShut {NoStop}%
\bibitem [{\citenamefont {Dalton}\ \emph {et~al.}(2013)\citenamefont {Dalton},
  \citenamefont {Hsieh}, \citenamefont {Hohensee}, \citenamefont {Cahill},\
  and\ \citenamefont {Goncharov}}]{ref:dalton2013}%
  \BibitemOpen
  \bibfield  {author} {\bibinfo {author} {\bibfnamefont {D.~A.}\ \bibnamefont
  {Dalton}}, \bibinfo {author} {\bibfnamefont {W.~P.}\ \bibnamefont {Hsieh}},
  \bibinfo {author} {\bibfnamefont {G.~T.}\ \bibnamefont {Hohensee}}, \bibinfo
  {author} {\bibfnamefont {D.~G.}\ \bibnamefont {Cahill}},\ and\ \bibinfo
  {author} {\bibfnamefont {A.~F.}\ \bibnamefont {Goncharov}},\ }\bibfield
  {title} {\bibinfo {title} {Effect of mass disorder on the lattice thermal
  conductivity of mgo periclase under pressure},\ }\href
  {https://doi.org/10.1038/srep02400} {\bibfield  {journal} {\bibinfo
  {journal} {Sci Rep}\ }\textbf {\bibinfo {volume} {3}},\ \bibinfo {pages}
  {2400} (\bibinfo {year} {2013})}\BibitemShut {NoStop}%
\bibitem [{\citenamefont {Hasegawa}\ \emph {et~al.}(2023)\citenamefont
  {Hasegawa}, \citenamefont {Ohta}, \citenamefont {Yagi},\ and\ \citenamefont
  {Hirose}}]{ref:hasegawa2023}%
  \BibitemOpen
  \bibfield  {author} {\bibinfo {author} {\bibfnamefont {A.}~\bibnamefont
  {Hasegawa}}, \bibinfo {author} {\bibfnamefont {K.}~\bibnamefont {Ohta}},
  \bibinfo {author} {\bibfnamefont {T.}~\bibnamefont {Yagi}},\ and\ \bibinfo
  {author} {\bibfnamefont {K.}~\bibnamefont {Hirose}},\ }\bibfield  {title}
  {\bibinfo {title} {Thermal conductivity of platinum and periclase under
  extreme conditions of pressure and temperature},\ }\href
  {https://doi.org/10.1080/08957959.2023.2193892} {\bibfield  {journal}
  {\bibinfo  {journal} {High Pressure Research}\ }\textbf {\bibinfo {volume}
  {43}},\ \bibinfo {pages} {68} (\bibinfo {year} {2023})}\BibitemShut {NoStop}%
\bibitem [{\citenamefont {Stackhouse}\ \emph {et~al.}(2010)\citenamefont
  {Stackhouse}, \citenamefont {Stixrude},\ and\ \citenamefont
  {Karki}}]{ref:stackhouse2010}%
  \BibitemOpen
  \bibfield  {author} {\bibinfo {author} {\bibfnamefont {S.}~\bibnamefont
  {Stackhouse}}, \bibinfo {author} {\bibfnamefont {L.}~\bibnamefont
  {Stixrude}},\ and\ \bibinfo {author} {\bibfnamefont {B.~B.}\ \bibnamefont
  {Karki}},\ }\bibfield  {title} {\bibinfo {title} {Thermal conductivity of
  periclase (mgo) from first principles},\ }\href
  {https://doi.org/10.1103/PhysRevLett.104.208501} {\bibfield  {journal}
  {\bibinfo  {journal} {Phys Rev Lett}\ }\textbf {\bibinfo {volume} {104}},\
  \bibinfo {pages} {208501} (\bibinfo {year} {2010})}\BibitemShut {NoStop}%
\bibitem [{\citenamefont {Haigis}\ \emph {et~al.}(2012)\citenamefont {Haigis},
  \citenamefont {Salanne},\ and\ \citenamefont {Jahn}}]{ref:haigis2012}%
  \BibitemOpen
  \bibfield  {author} {\bibinfo {author} {\bibfnamefont {V.}~\bibnamefont
  {Haigis}}, \bibinfo {author} {\bibfnamefont {M.}~\bibnamefont {Salanne}},\
  and\ \bibinfo {author} {\bibfnamefont {S.}~\bibnamefont {Jahn}},\ }\bibfield
  {title} {\bibinfo {title} {Thermal conductivity of mgo, mgsio3 perovskite and
  post-perovskite in the earth's deep mantle},\ }\href
  {https://doi.org/10.1016/j.epsl.2012.09.002} {\bibfield  {journal} {\bibinfo
  {journal} {Earth and Planetary Science Letters}\ }\textbf {\bibinfo {volume}
  {355-356}},\ \bibinfo {pages} {102} (\bibinfo {year} {2012})}\BibitemShut
  {NoStop}%
\bibitem [{\citenamefont {Tang}\ and\ \citenamefont
  {Dong}(2010)}]{ref:tang2010}%
  \BibitemOpen
  \bibfield  {author} {\bibinfo {author} {\bibfnamefont {X.}~\bibnamefont
  {Tang}}\ and\ \bibinfo {author} {\bibfnamefont {J.}~\bibnamefont {Dong}},\
  }\bibfield  {title} {\bibinfo {title} {Lattice thermal conductivity of mgo at
  conditions of earth's interior},\ }\href
  {https://doi.org/10.1073/pnas.0907194107} {\bibfield  {journal} {\bibinfo
  {journal} {Proc Natl Acad Sci U S A}\ }\textbf {\bibinfo {volume} {107}},\
  \bibinfo {pages} {4539} (\bibinfo {year} {2010})}\BibitemShut {NoStop}%
\bibitem [{\citenamefont {Kwon}\ \emph {et~al.}(2020)\citenamefont {Kwon},
  \citenamefont {Xia}, \citenamefont {Zhou},\ and\ \citenamefont
  {Han}}]{ref:kwon2020}%
  \BibitemOpen
  \bibfield  {author} {\bibinfo {author} {\bibfnamefont {C.}~\bibnamefont
  {Kwon}}, \bibinfo {author} {\bibfnamefont {Y.}~\bibnamefont {Xia}}, \bibinfo
  {author} {\bibfnamefont {F.}~\bibnamefont {Zhou}},\ and\ \bibinfo {author}
  {\bibfnamefont {B.}~\bibnamefont {Han}},\ }\bibfield  {title} {\bibinfo
  {title} {Dominant effect of anharmonicity on the equation of state and
  thermal conductivity of mgo under extreme conditions},\ }\href
  {https://doi.org/10.1103/PhysRevB.102.184309} {\bibfield  {journal} {\bibinfo
   {journal} {Phys. Rev. B}\ }\textbf {\bibinfo {volume} {102}},\ \bibinfo
  {pages} {184309} (\bibinfo {year} {2020})}\BibitemShut {NoStop}%
\bibitem [{\citenamefont {Koker}(2010)}]{ref:dekoker2010}%
  \BibitemOpen
  \bibfield  {author} {\bibinfo {author} {\bibfnamefont {N.~d.}\ \bibnamefont
  {Koker}},\ }\bibfield  {title} {\bibinfo {title} {Thermal conductivity of mgo
  periclase at high pressure: Implications for the d'' region},\ }\href
  {https://doi.org/10.1016/j.epsl.2010.02.011} {\bibfield  {journal} {\bibinfo
  {journal} {Earth and Planetary Science Letters}\ }\textbf {\bibinfo {volume}
  {292}},\ \bibinfo {pages} {392} (\bibinfo {year} {2010})}\BibitemShut
  {NoStop}%
\bibitem [{\citenamefont {Ravichandran}\ and\ \citenamefont
  {Broido}(2019)}]{ref:Ravichandran2019}%
  \BibitemOpen
  \bibfield  {author} {\bibinfo {author} {\bibfnamefont {N.~K.}\ \bibnamefont
  {Ravichandran}}\ and\ \bibinfo {author} {\bibfnamefont {D.}~\bibnamefont
  {Broido}},\ }\bibfield  {title} {\bibinfo {title} {Non-monotonic pressure
  dependence of the thermal conductivity of boron arsenide},\ }\href
  {https://doi.org/10.1038/s41467-019-08713-0} {\bibfield  {journal} {\bibinfo
  {journal} {Nature Communications}\ }\textbf {\bibinfo {volume} {10}},\
  \bibinfo {pages} {827} (\bibinfo {year} {2019})}\BibitemShut {NoStop}%
\bibitem [{\citenamefont {Soubiran}\ and\ \citenamefont
  {Militzer}(2020)}]{ref:soubiran2020}%
  \BibitemOpen
  \bibfield  {author} {\bibinfo {author} {\bibfnamefont {F.}~\bibnamefont
  {Soubiran}}\ and\ \bibinfo {author} {\bibfnamefont {B.}~\bibnamefont
  {Militzer}},\ }\bibfield  {title} {\bibinfo {title} {Anharmonicity and phase
  diagram of magnesium oxide in the megabar regime},\ }\href
  {https://doi.org/10.1103/PhysRevLett.125.175701} {\bibfield  {journal}
  {\bibinfo  {journal} {Phys Rev Lett}\ }\textbf {\bibinfo {volume} {125}},\
  \bibinfo {pages} {175701} (\bibinfo {year} {2020})}\BibitemShut {NoStop}%
\bibitem [{\citenamefont {Zhang}\ \emph {et~al.}(2018)\citenamefont {Zhang},
  \citenamefont {Han}, \citenamefont {Wang}, \citenamefont {Car},\ and\
  \citenamefont {Weinan}}]{ref:zhang2018deep}%
  \BibitemOpen
  \bibfield  {author} {\bibinfo {author} {\bibfnamefont {L.}~\bibnamefont
  {Zhang}}, \bibinfo {author} {\bibfnamefont {J.}~\bibnamefont {Han}}, \bibinfo
  {author} {\bibfnamefont {H.}~\bibnamefont {Wang}}, \bibinfo {author}
  {\bibfnamefont {R.}~\bibnamefont {Car}},\ and\ \bibinfo {author}
  {\bibfnamefont {E.}~\bibnamefont {Weinan}},\ }\bibfield  {title} {\bibinfo
  {title} {Deep potential molecular dynamics: a scalable model with the
  accuracy of quantum mechanics},\ }\href
  {https://doi.org/10.1103/PhysRevLett.120.143001} {\bibfield  {journal}
  {\bibinfo  {journal} {Phys. Rev. Lett.}\ }\textbf {\bibinfo {volume} {120}},\
  \bibinfo {pages} {143001} (\bibinfo {year} {2018})}\BibitemShut {NoStop}%
\bibitem [{\citenamefont {Zeng}\ \emph {et~al.}(2021)\citenamefont {Zeng},
  \citenamefont {Yu}, \citenamefont {Yao}, \citenamefont {Gao}, \citenamefont
  {Chen}, \citenamefont {Zhang}, \citenamefont {Kang}, \citenamefont {Wang},\
  and\ \citenamefont {Dai}}]{ref:zeng2021ab}%
  \BibitemOpen
  \bibfield  {author} {\bibinfo {author} {\bibfnamefont {Q.}~\bibnamefont
  {Zeng}}, \bibinfo {author} {\bibfnamefont {X.}~\bibnamefont {Yu}}, \bibinfo
  {author} {\bibfnamefont {Y.}~\bibnamefont {Yao}}, \bibinfo {author}
  {\bibfnamefont {T.}~\bibnamefont {Gao}}, \bibinfo {author} {\bibfnamefont
  {B.}~\bibnamefont {Chen}}, \bibinfo {author} {\bibfnamefont {S.}~\bibnamefont
  {Zhang}}, \bibinfo {author} {\bibfnamefont {D.}~\bibnamefont {Kang}},
  \bibinfo {author} {\bibfnamefont {H.}~\bibnamefont {Wang}},\ and\ \bibinfo
  {author} {\bibfnamefont {J.}~\bibnamefont {Dai}},\ }\bibfield  {title}
  {\bibinfo {title} {Ab initio validation on the connection between atomistic
  and hydrodynamic description to unravel the ion dynamics of warm dense
  matter},\ }\href {https://doi.org/10.1103/PhysRevResearch.3.033116}
  {\bibfield  {journal} {\bibinfo  {journal} {Phys. Rev. Research}\ }\textbf
  {\bibinfo {volume} {3}},\ \bibinfo {pages} {033116} (\bibinfo {year}
  {2021})}\BibitemShut {NoStop}%
\bibitem [{\citenamefont {Zeng}\ \emph {et~al.}(2022)\citenamefont {Zeng},
  \citenamefont {Chen}, \citenamefont {Yu}, \citenamefont {Zhang},
  \citenamefont {Kang}, \citenamefont {Wang},\ and\ \citenamefont
  {Dai}}]{ref:zeng2022towards}%
  \BibitemOpen
  \bibfield  {author} {\bibinfo {author} {\bibfnamefont {Q.}~\bibnamefont
  {Zeng}}, \bibinfo {author} {\bibfnamefont {B.}~\bibnamefont {Chen}}, \bibinfo
  {author} {\bibfnamefont {X.}~\bibnamefont {Yu}}, \bibinfo {author}
  {\bibfnamefont {S.}~\bibnamefont {Zhang}}, \bibinfo {author} {\bibfnamefont
  {D.}~\bibnamefont {Kang}}, \bibinfo {author} {\bibfnamefont {H.}~\bibnamefont
  {Wang}},\ and\ \bibinfo {author} {\bibfnamefont {J.}~\bibnamefont {Dai}},\
  }\bibfield  {title} {\bibinfo {title} {Towards large-scale and
  spatiotemporally resolved diagnosis of electronic density of states by deep
  learning},\ }\href {https://doi.org/10.1103/PhysRevB.105.174109} {\bibfield
  {journal} {\bibinfo  {journal} {Phys. Rev. B}\ }\textbf {\bibinfo {volume}
  {105}},\ \bibinfo {pages} {174109} (\bibinfo {year} {2022})}\BibitemShut
  {NoStop}%
\bibitem [{\citenamefont {Deng}\ and\ \citenamefont
  {Stixrude}(2021)}]{ref:Deng2021}%
  \BibitemOpen
  \bibfield  {author} {\bibinfo {author} {\bibfnamefont {J.}~\bibnamefont
  {Deng}}\ and\ \bibinfo {author} {\bibfnamefont {L.}~\bibnamefont
  {Stixrude}},\ }\bibfield  {title} {\bibinfo {title} {Thermal conductivity of
  silicate liquid determined by machine learning potentials},\ }\href
  {https://doi.org/https://doi.org/10.1029/2021GL093806} {\bibfield  {journal}
  {\bibinfo  {journal} {Geophysical Research Letters}\ }\textbf {\bibinfo
  {volume} {48}},\ \bibinfo {pages} {e2021GL093806} (\bibinfo {year}
  {2021})}\BibitemShut {NoStop}%
\bibitem [{\citenamefont {Liu}\ \emph {et~al.}(2021)\citenamefont {Liu},
  \citenamefont {Li},\ and\ \citenamefont {Chen}}]{ref:Liu2021}%
  \BibitemOpen
  \bibfield  {author} {\bibinfo {author} {\bibfnamefont {Q.}~\bibnamefont
  {Liu}}, \bibinfo {author} {\bibfnamefont {J.}~\bibnamefont {Li}},\ and\
  \bibinfo {author} {\bibfnamefont {M.}~\bibnamefont {Chen}},\ }\bibfield
  {title} {\bibinfo {title} {Thermal transport by electrons and ions in warm
  dense aluminum: A combined density functional theory and deep potential
  study},\ }\href {https://doi.org/10.1063/5.0030123} {\bibfield  {journal}
  {\bibinfo  {journal} {Matter and Radiation at Extremes}\ }\textbf {\bibinfo
  {volume} {6}},\ \bibinfo {pages} {026902} (\bibinfo {year}
  {2021})}\BibitemShut {NoStop}%
\bibitem [{\citenamefont {Wang}\ \emph {et~al.}(2022)\citenamefont {Wang},
  \citenamefont {Wu},\ and\ \citenamefont {Deng}}]{ref:Wang2022}%
  \BibitemOpen
  \bibfield  {author} {\bibinfo {author} {\bibfnamefont {D.}~\bibnamefont
  {Wang}}, \bibinfo {author} {\bibfnamefont {Z.}~\bibnamefont {Wu}},\ and\
  \bibinfo {author} {\bibfnamefont {X.}~\bibnamefont {Deng}},\ }\bibfield
  {title} {\bibinfo {title} {Thermal conductivity of hydrous wadsleyite
  determined by non-equilibrium molecular dynamics based on machine learning},\
  }\href {https://doi.org/https://doi.org/10.1029/2022GL100337} {\bibfield
  {journal} {\bibinfo  {journal} {Geophysical Research Letters}\ }\textbf
  {\bibinfo {volume} {49}},\ \bibinfo {pages} {e2022GL100337} (\bibinfo {year}
  {2022})}\BibitemShut {NoStop}%
\bibitem [{\citenamefont {Yang}\ \emph {et~al.}(2022)\citenamefont {Yang},
  \citenamefont {Zeng}, \citenamefont {Chen}, \citenamefont {Kang},
  \citenamefont {Zhang}, \citenamefont {Wu}, \citenamefont {Yu},\ and\
  \citenamefont {Dai}}]{ref:Yang_2022}%
  \BibitemOpen
  \bibfield  {author} {\bibinfo {author} {\bibfnamefont {F.}~\bibnamefont
  {Yang}}, \bibinfo {author} {\bibfnamefont {Q.}~\bibnamefont {Zeng}}, \bibinfo
  {author} {\bibfnamefont {B.}~\bibnamefont {Chen}}, \bibinfo {author}
  {\bibfnamefont {D.}~\bibnamefont {Kang}}, \bibinfo {author} {\bibfnamefont
  {S.}~\bibnamefont {Zhang}}, \bibinfo {author} {\bibfnamefont
  {J.}~\bibnamefont {Wu}}, \bibinfo {author} {\bibfnamefont {X.}~\bibnamefont
  {Yu}},\ and\ \bibinfo {author} {\bibfnamefont {J.}~\bibnamefont {Dai}},\
  }\bibfield  {title} {\bibinfo {title} {Lattice thermal conductivity of mgsio3
  perovskite and post-perovskite under lower mantle conditions calculated by
  deep potential molecular dynamics},\ }\href
  {https://doi.org/10.1088/0256-307X/39/11/116301} {\bibfield  {journal}
  {\bibinfo  {journal} {Chinese Physics Letters}\ }\textbf {\bibinfo {volume}
  {39}},\ \bibinfo {pages} {116301} (\bibinfo {year} {2022})}\BibitemShut
  {NoStop}%
\bibitem [{\citenamefont {Zhang}\ \emph
  {et~al.}(2023{\natexlab{a}})\citenamefont {Zhang}, \citenamefont
  {Puligheddu}, \citenamefont {Zhang}, \citenamefont {Car},\ and\ \citenamefont
  {Galli}}]{ref:Zhang2023thermal}%
  \BibitemOpen
  \bibfield  {author} {\bibinfo {author} {\bibfnamefont {C.}~\bibnamefont
  {Zhang}}, \bibinfo {author} {\bibfnamefont {M.}~\bibnamefont {Puligheddu}},
  \bibinfo {author} {\bibfnamefont {L.}~\bibnamefont {Zhang}}, \bibinfo
  {author} {\bibfnamefont {R.}~\bibnamefont {Car}},\ and\ \bibinfo {author}
  {\bibfnamefont {G.}~\bibnamefont {Galli}},\ }\bibfield  {title} {\bibinfo
  {title} {Thermal conductivity of water at extreme conditions},\ }\href
  {https://doi.org/10.1021/acs.jpcb.3c02972} {\bibfield  {journal} {\bibinfo
  {journal} {The Journal of Physical Chemistry B}\ }\textbf {\bibinfo {volume}
  {127}},\ \bibinfo {pages} {7011} (\bibinfo {year} {2023}{\natexlab{a}})},\
  \bibinfo {note} {pMID: 37524047}\BibitemShut {NoStop}%
\bibitem [{\citenamefont {Zeng}\ \emph {et~al.}(2023)\citenamefont {Zeng},
  \citenamefont {Zhang}, \citenamefont {Lu}, \citenamefont {Mo}, \citenamefont
  {Li}, \citenamefont {Chen}, \citenamefont {Rynik}, \citenamefont {Huang},
  \citenamefont {Li}, \citenamefont {Shi}, \citenamefont {Wang}, \citenamefont
  {Ye}, \citenamefont {Tuo}, \citenamefont {Yang}, \citenamefont {Ding},
  \citenamefont {Li}, \citenamefont {Tisi}, \citenamefont {Zeng}, \citenamefont
  {Bao}, \citenamefont {Xia}, \citenamefont {Huang}, \citenamefont {Muraoka},
  \citenamefont {Wang}, \citenamefont {Chang}, \citenamefont {Yuan},
  \citenamefont {Bore}, \citenamefont {Cai}, \citenamefont {Lin}, \citenamefont
  {Wang}, \citenamefont {Xu}, \citenamefont {Zhu}, \citenamefont {Luo},
  \citenamefont {Zhang}, \citenamefont {Goodall}, \citenamefont {Liang},
  \citenamefont {Singh}, \citenamefont {Yao}, \citenamefont {Zhang},
  \citenamefont {Wentzcovitch}, \citenamefont {Han}, \citenamefont {Liu},
  \citenamefont {Jia}, \citenamefont {York}, \citenamefont {E}, \citenamefont
  {Car}, \citenamefont {Zhang},\ and\ \citenamefont {Wang}}]{dpkitv2}%
  \BibitemOpen
  \bibfield  {author} {\bibinfo {author} {\bibfnamefont {J.}~\bibnamefont
  {Zeng}}, \bibinfo {author} {\bibfnamefont {D.}~\bibnamefont {Zhang}},
  \bibinfo {author} {\bibfnamefont {D.}~\bibnamefont {Lu}}, \bibinfo {author}
  {\bibfnamefont {P.}~\bibnamefont {Mo}}, \bibinfo {author} {\bibfnamefont
  {Z.}~\bibnamefont {Li}}, \bibinfo {author} {\bibfnamefont {Y.}~\bibnamefont
  {Chen}}, \bibinfo {author} {\bibfnamefont {M.}~\bibnamefont {Rynik}},
  \bibinfo {author} {\bibfnamefont {L.}~\bibnamefont {Huang}}, \bibinfo
  {author} {\bibfnamefont {Z.}~\bibnamefont {Li}}, \bibinfo {author}
  {\bibfnamefont {S.}~\bibnamefont {Shi}}, \bibinfo {author} {\bibfnamefont
  {Y.}~\bibnamefont {Wang}}, \bibinfo {author} {\bibfnamefont {H.}~\bibnamefont
  {Ye}}, \bibinfo {author} {\bibfnamefont {P.}~\bibnamefont {Tuo}}, \bibinfo
  {author} {\bibfnamefont {J.}~\bibnamefont {Yang}}, \bibinfo {author}
  {\bibfnamefont {Y.}~\bibnamefont {Ding}}, \bibinfo {author} {\bibfnamefont
  {Y.}~\bibnamefont {Li}}, \bibinfo {author} {\bibfnamefont {D.}~\bibnamefont
  {Tisi}}, \bibinfo {author} {\bibfnamefont {Q.}~\bibnamefont {Zeng}}, \bibinfo
  {author} {\bibfnamefont {H.}~\bibnamefont {Bao}}, \bibinfo {author}
  {\bibfnamefont {Y.}~\bibnamefont {Xia}}, \bibinfo {author} {\bibfnamefont
  {J.}~\bibnamefont {Huang}}, \bibinfo {author} {\bibfnamefont
  {K.}~\bibnamefont {Muraoka}}, \bibinfo {author} {\bibfnamefont
  {Y.}~\bibnamefont {Wang}}, \bibinfo {author} {\bibfnamefont {J.}~\bibnamefont
  {Chang}}, \bibinfo {author} {\bibfnamefont {F.}~\bibnamefont {Yuan}},
  \bibinfo {author} {\bibfnamefont {S.~L.}\ \bibnamefont {Bore}}, \bibinfo
  {author} {\bibfnamefont {C.}~\bibnamefont {Cai}}, \bibinfo {author}
  {\bibfnamefont {Y.}~\bibnamefont {Lin}}, \bibinfo {author} {\bibfnamefont
  {B.}~\bibnamefont {Wang}}, \bibinfo {author} {\bibfnamefont {J.}~\bibnamefont
  {Xu}}, \bibinfo {author} {\bibfnamefont {J.~X.}\ \bibnamefont {Zhu}},
  \bibinfo {author} {\bibfnamefont {C.}~\bibnamefont {Luo}}, \bibinfo {author}
  {\bibfnamefont {Y.}~\bibnamefont {Zhang}}, \bibinfo {author} {\bibfnamefont
  {R.~E.~A.}\ \bibnamefont {Goodall}}, \bibinfo {author} {\bibfnamefont
  {W.}~\bibnamefont {Liang}}, \bibinfo {author} {\bibfnamefont {A.~K.}\
  \bibnamefont {Singh}}, \bibinfo {author} {\bibfnamefont {S.}~\bibnamefont
  {Yao}}, \bibinfo {author} {\bibfnamefont {J.}~\bibnamefont {Zhang}}, \bibinfo
  {author} {\bibfnamefont {R.}~\bibnamefont {Wentzcovitch}}, \bibinfo {author}
  {\bibfnamefont {J.}~\bibnamefont {Han}}, \bibinfo {author} {\bibfnamefont
  {J.}~\bibnamefont {Liu}}, \bibinfo {author} {\bibfnamefont {W.}~\bibnamefont
  {Jia}}, \bibinfo {author} {\bibfnamefont {D.~M.}\ \bibnamefont {York}},
  \bibinfo {author} {\bibfnamefont {W.}~\bibnamefont {E}}, \bibinfo {author}
  {\bibfnamefont {R.}~\bibnamefont {Car}}, \bibinfo {author} {\bibfnamefont
  {L.}~\bibnamefont {Zhang}},\ and\ \bibinfo {author} {\bibfnamefont
  {H.}~\bibnamefont {Wang}},\ }\bibfield  {title} {\bibinfo {title} {Deepmd-kit
  v2: A software package for deep potential models},\ }\href
  {https://doi.org/10.1063/5.0155600} {\bibfield  {journal} {\bibinfo
  {journal} {J Chem Phys}\ }\textbf {\bibinfo {volume} {159}},\ \bibinfo
  {pages} {054801} (\bibinfo {year} {2023})}\BibitemShut {NoStop}%
\bibitem [{\citenamefont {Zhang}\ \emph {et~al.}(2019)\citenamefont {Zhang},
  \citenamefont {Lin}, \citenamefont {Wang}, \citenamefont {Car},\ and\
  \citenamefont {Weinan}}]{ref:zhang2019active}%
  \BibitemOpen
  \bibfield  {author} {\bibinfo {author} {\bibfnamefont {L.}~\bibnamefont
  {Zhang}}, \bibinfo {author} {\bibfnamefont {D.-Y.}\ \bibnamefont {Lin}},
  \bibinfo {author} {\bibfnamefont {H.}~\bibnamefont {Wang}}, \bibinfo {author}
  {\bibfnamefont {R.}~\bibnamefont {Car}},\ and\ \bibinfo {author}
  {\bibfnamefont {E.}~\bibnamefont {Weinan}},\ }\bibfield  {title} {\bibinfo
  {title} {Active learning of uniformly accurate interatomic potentials for
  materials simulation},\ }\href
  {https://doi.org/10.1103/PhysRevMaterials.3.023804} {\bibfield  {journal}
  {\bibinfo  {journal} {Phys. Rev. Materials}\ }\textbf {\bibinfo {volume}
  {3}},\ \bibinfo {pages} {023804} (\bibinfo {year} {2019})}\BibitemShut
  {NoStop}%
\bibitem [{\citenamefont {Kresse}\ and\ \citenamefont
  {Furthm\"uller}(1996{\natexlab{a}})}]{ref:KRESSE199615_vasp}%
  \BibitemOpen
  \bibfield  {author} {\bibinfo {author} {\bibfnamefont {G.}~\bibnamefont
  {Kresse}}\ and\ \bibinfo {author} {\bibfnamefont {J.}~\bibnamefont
  {Furthm\"uller}},\ }\bibfield  {title} {\bibinfo {title} {Efficiency of
  ab-initio total energy calculations for metals and semiconductors using a
  plane-wave basis set},\ }\href
  {https://doi.org/https://doi.org/10.1016/0927-0256(96)00008-0} {\bibfield
  {journal} {\bibinfo  {journal} {Computational Materials Science}\ }\textbf
  {\bibinfo {volume} {6}},\ \bibinfo {pages} {15} (\bibinfo {year}
  {1996}{\natexlab{a}})}\BibitemShut {NoStop}%
\bibitem [{\citenamefont {Kresse}\ and\ \citenamefont
  {Furthm\"uller}(1996{\natexlab{b}})}]{ref:Kress_vasp}%
  \BibitemOpen
  \bibfield  {author} {\bibinfo {author} {\bibfnamefont {G.}~\bibnamefont
  {Kresse}}\ and\ \bibinfo {author} {\bibfnamefont {J.}~\bibnamefont
  {Furthm\"uller}},\ }\bibfield  {title} {\bibinfo {title} {Efficient iterative
  schemes for ab initio total-energy calculations using a plane-wave basis
  set},\ }\href {https://doi.org/10.1103/PhysRevB.54.11169} {\bibfield
  {journal} {\bibinfo  {journal} {Phys. Rev. B}\ }\textbf {\bibinfo {volume}
  {54}},\ \bibinfo {pages} {11169} (\bibinfo {year}
  {1996}{\natexlab{b}})}\BibitemShut {NoStop}%
\bibitem [{\citenamefont {Perdew}\ \emph {et~al.}(1996)\citenamefont {Perdew},
  \citenamefont {Burke},\ and\ \citenamefont {Ernzerhof}}]{ref:perdew_1996}%
  \BibitemOpen
  \bibfield  {author} {\bibinfo {author} {\bibfnamefont {J.~P.}\ \bibnamefont
  {Perdew}}, \bibinfo {author} {\bibfnamefont {K.}~\bibnamefont {Burke}},\ and\
  \bibinfo {author} {\bibfnamefont {M.}~\bibnamefont {Ernzerhof}},\ }\bibfield
  {title} {\bibinfo {title} {Generalized gradient approximation made simple},\
  }\href {https://doi.org/10.1103/PhysRevLett.77.3865} {\bibfield  {journal}
  {\bibinfo  {journal} {Phys. Rev. Lett.}\ }\textbf {\bibinfo {volume} {77}},\
  \bibinfo {pages} {3865} (\bibinfo {year} {1996})}\BibitemShut {NoStop}%
\bibitem [{\citenamefont {Perdew}\ \emph {et~al.}(2008)\citenamefont {Perdew},
  \citenamefont {Ruzsinszky}, \citenamefont {Csonka}, \citenamefont {Vydrov},
  \citenamefont {Scuseria}, \citenamefont {Constantin}, \citenamefont {Zhou},\
  and\ \citenamefont {Burke}}]{ref:perdew2008}%
  \BibitemOpen
  \bibfield  {author} {\bibinfo {author} {\bibfnamefont {J.~P.}\ \bibnamefont
  {Perdew}}, \bibinfo {author} {\bibfnamefont {A.}~\bibnamefont {Ruzsinszky}},
  \bibinfo {author} {\bibfnamefont {G.~I.}\ \bibnamefont {Csonka}}, \bibinfo
  {author} {\bibfnamefont {O.~A.}\ \bibnamefont {Vydrov}}, \bibinfo {author}
  {\bibfnamefont {G.~E.}\ \bibnamefont {Scuseria}}, \bibinfo {author}
  {\bibfnamefont {L.~A.}\ \bibnamefont {Constantin}}, \bibinfo {author}
  {\bibfnamefont {X.}~\bibnamefont {Zhou}},\ and\ \bibinfo {author}
  {\bibfnamefont {K.}~\bibnamefont {Burke}},\ }\bibfield  {title} {\bibinfo
  {title} {Restoring the density-gradient expansion for exchange in solids and
  surfaces},\ }\bibfield  {journal} {\bibinfo  {journal} {Physical Review
  Letters}\ }\textbf {\bibinfo {volume} {100}},\ \href
  {https://doi.org/10.1103/PhysRevLett.100.136406}
  {10.1103/PhysRevLett.100.136406} (\bibinfo {year} {2008})\BibitemShut
  {NoStop}%
\bibitem [{\citenamefont {Zhang}\ \emph
  {et~al.}(2023{\natexlab{b}})\citenamefont {Zhang}, \citenamefont {Paul},
  \citenamefont {Hu},\ and\ \citenamefont {Morales}}]{ref:zhang2023}%
  \BibitemOpen
  \bibfield  {author} {\bibinfo {author} {\bibfnamefont {S.}~\bibnamefont
  {Zhang}}, \bibinfo {author} {\bibfnamefont {R.}~\bibnamefont {Paul}},
  \bibinfo {author} {\bibfnamefont {S.~X.}\ \bibnamefont {Hu}},\ and\ \bibinfo
  {author} {\bibfnamefont {M.~A.}\ \bibnamefont {Morales}},\ }\bibfield
  {title} {\bibinfo {title} {Toward an accurate equation?of state and b1-b2
  phase boundary for magnesium oxide up to terapascal pressures and
  electron-volt temperatures},\ }\href
  {https://doi.org/10.1103/PhysRevB.107.224109} {\bibfield  {journal} {\bibinfo
   {journal} {Physical Review B}\ }\textbf {\bibinfo {volume} {107}},\ \bibinfo
  {pages} {4109} (\bibinfo {year} {2023}{\natexlab{b}})}\BibitemShut {NoStop}%
\bibitem [{\citenamefont {Bl{\"o}chl}(1994)}]{ref:blochl1994projector}%
  \BibitemOpen
  \bibfield  {author} {\bibinfo {author} {\bibfnamefont {P.~E.}\ \bibnamefont
  {Bl{\"o}chl}},\ }\bibfield  {title} {\bibinfo {title} {Projector
  augmented-wave method},\ }\href {https://doi.org/10.1103/PhysRevB.50.17953}
  {\bibfield  {journal} {\bibinfo  {journal} {Phys. Rev. B}\ }\textbf {\bibinfo
  {volume} {50}},\ \bibinfo {pages} {17953} (\bibinfo {year}
  {1994})}\BibitemShut {NoStop}%
\bibitem [{\citenamefont {Holzwarth}\ \emph {et~al.}(2001)\citenamefont
  {Holzwarth}, \citenamefont {Tackett},\ and\ \citenamefont
  {Matthews}}]{ref:holzwarth2001projector}%
  \BibitemOpen
  \bibfield  {author} {\bibinfo {author} {\bibfnamefont {N.}~\bibnamefont
  {Holzwarth}}, \bibinfo {author} {\bibfnamefont {A.}~\bibnamefont {Tackett}},\
  and\ \bibinfo {author} {\bibfnamefont {G.}~\bibnamefont {Matthews}},\
  }\bibfield  {title} {\bibinfo {title} {A projector augmented wave (paw) code
  for electronic structure calculations, part i: atompaw for generating
  atom-centered functions},\ }\href
  {https://doi.org/10.1016/S0010-4655(00)00244-7} {\bibfield  {journal}
  {\bibinfo  {journal} {Computer Physics Communications}\ }\textbf {\bibinfo
  {volume} {135}},\ \bibinfo {pages} {329} (\bibinfo {year}
  {2001})}\BibitemShut {NoStop}%
\bibitem [{\citenamefont {Mcquarrie}(1965)}]{ref:mcquarrie1965statistical}%
  \BibitemOpen
  \bibfield  {author} {\bibinfo {author} {\bibfnamefont {D.}~\bibnamefont
  {Mcquarrie}},\ }\href@noop {} {\bibinfo {title} {Statistical mechanics}}
  (\bibinfo {year} {1965})\BibitemShut {NoStop}%
\bibitem [{\citenamefont {Plimpton}(1995)}]{ref:plimpton1995fast}%
  \BibitemOpen
  \bibfield  {author} {\bibinfo {author} {\bibfnamefont {S.}~\bibnamefont
  {Plimpton}},\ }\bibfield  {title} {\bibinfo {title} {Fast parallel algorithms
  for short-range molecular dynamics},\ }\href
  {https://doi.org/10.1006/jcph.1995.1039} {\bibfield  {journal} {\bibinfo
  {journal} {Journal of computational physics}\ }\textbf {\bibinfo {volume}
  {117}},\ \bibinfo {pages} {1} (\bibinfo {year} {1995})}\BibitemShut {NoStop}%
\bibitem [{\citenamefont {Nos{\'e}}(1984)}]{ref:nose1984unified}%
  \BibitemOpen
  \bibfield  {author} {\bibinfo {author} {\bibfnamefont {S.}~\bibnamefont
  {Nos{\'e}}},\ }\bibfield  {title} {\bibinfo {title} {A unified formulation of
  the constant temperature molecular dynamics methods},\ }\href
  {https://doi.org/10.1063/1.447334} {\bibfield  {journal} {\bibinfo  {journal}
  {The Journal of chemical physics}\ }\textbf {\bibinfo {volume} {81}},\
  \bibinfo {pages} {511} (\bibinfo {year} {1984})}\BibitemShut {NoStop}%
\bibitem [{\citenamefont {Hoover}(1985)}]{ref:hoover1985canonical}%
  \BibitemOpen
  \bibfield  {author} {\bibinfo {author} {\bibfnamefont {W.~G.}\ \bibnamefont
  {Hoover}},\ }\bibfield  {title} {\bibinfo {title} {Canonical dynamics:
  Equilibrium phase-space distributions},\ }\href
  {https://doi.org/10.1103/PhysRevA.31.1695} {\bibfield  {journal} {\bibinfo
  {journal} {Phys. Rev. A}\ }\textbf {\bibinfo {volume} {31}},\ \bibinfo
  {pages} {1695} (\bibinfo {year} {1985})}\BibitemShut {NoStop}%
\bibitem [{\citenamefont {Tadano}\ \emph {et~al.}(2014)\citenamefont {Tadano},
  \citenamefont {Gohda},\ and\ \citenamefont {Tsuneyuki}}]{ref:tadano2014}%
  \BibitemOpen
  \bibfield  {author} {\bibinfo {author} {\bibfnamefont {T.}~\bibnamefont
  {Tadano}}, \bibinfo {author} {\bibfnamefont {Y.}~\bibnamefont {Gohda}},\ and\
  \bibinfo {author} {\bibfnamefont {S.}~\bibnamefont {Tsuneyuki}},\ }\bibfield
  {title} {\bibinfo {title} {Anharmonic force constants extracted from
  first-principles molecular dynamics: applications to heat transfer
  simulations},\ }\href {https://doi.org/10.1088/0953-8984/26/22/225402}
  {\bibfield  {journal} {\bibinfo  {journal} {J Phys Condens Matter}\ }\textbf
  {\bibinfo {volume} {26}},\ \bibinfo {pages} {225402} (\bibinfo {year}
  {2014})}\BibitemShut {NoStop}%
\bibitem [{\citenamefont {Sangster}\ \emph {et~al.}(1969)\citenamefont
  {Sangster}, \citenamefont {Peckham},\ and\ \citenamefont
  {Saunderson}}]{ref:Sangster1969}%
  \BibitemOpen
  \bibfield  {author} {\bibinfo {author} {\bibfnamefont {M.~J.~L.}\
  \bibnamefont {Sangster}}, \bibinfo {author} {\bibfnamefont {G.}~\bibnamefont
  {Peckham}},\ and\ \bibinfo {author} {\bibfnamefont {D.~H.}\ \bibnamefont
  {Saunderson}},\ }\bibfield  {title} {\bibinfo {title} {Lattice dynamics of
  magnesium oxide},\ }\href {https://doi.org/10.1088/0022-3719/3/5/017}
  {\bibfield  {journal} {\bibinfo  {journal} {J. PHYS. C: SOLID ST. PHYS.}\
  }\textbf {\bibinfo {volume} {3}},\ \bibinfo {pages} {1026} (\bibinfo {year}
  {1969})}\BibitemShut {NoStop}%
\bibitem [{\citenamefont {Hofmeister}(2014)}]{ref:hofmeister2014}%
  \BibitemOpen
  \bibfield  {author} {\bibinfo {author} {\bibfnamefont {A.~M.}\ \bibnamefont
  {Hofmeister}},\ }\bibfield  {title} {\bibinfo {title} {Thermal diffusivity
  and thermal conductivity of single-crystal mgo and al2o3 and related
  compounds as a function of temperature},\ }\href
  {https://doi.org/10.1007/s00269-014-0655-3} {\bibfield  {journal} {\bibinfo
  {journal} {Physics and Chemistry of Minerals}\ }\textbf {\bibinfo {volume}
  {41}},\ \bibinfo {pages} {361} (\bibinfo {year} {2014})}\BibitemShut
  {NoStop}%
\bibitem [{\citenamefont {Sellan}\ \emph {et~al.}(2010)\citenamefont {Sellan},
  \citenamefont {Landry}, \citenamefont {Turney}, \citenamefont {McGaughey},\
  and\ \citenamefont {Amon}}]{ref:Sellan2010}%
  \BibitemOpen
  \bibfield  {author} {\bibinfo {author} {\bibfnamefont {D.~P.}\ \bibnamefont
  {Sellan}}, \bibinfo {author} {\bibfnamefont {E.~S.}\ \bibnamefont {Landry}},
  \bibinfo {author} {\bibfnamefont {J.~E.}\ \bibnamefont {Turney}}, \bibinfo
  {author} {\bibfnamefont {A.~J.~H.}\ \bibnamefont {McGaughey}},\ and\ \bibinfo
  {author} {\bibfnamefont {C.~H.}\ \bibnamefont {Amon}},\ }\bibfield  {title}
  {\bibinfo {title} {Size effects in molecular dynamics thermal conductivity
  predictions},\ }\href {https://doi.org/10.1103/PhysRevB.81.214305} {\bibfield
   {journal} {\bibinfo  {journal} {Physical Review B}\ }\textbf {\bibinfo
  {volume} {81}},\ \bibinfo {pages} {214305} (\bibinfo {year}
  {2010})}\BibitemShut {NoStop}%
\bibitem [{\citenamefont {Dong}\ \emph {et~al.}(2018)\citenamefont {Dong},
  \citenamefont {Fan}, \citenamefont {Shi}, \citenamefont {Harju},\ and\
  \citenamefont {Ala-Nissila}}]{ref:Dong2018}%
  \BibitemOpen
  \bibfield  {author} {\bibinfo {author} {\bibfnamefont {H.}~\bibnamefont
  {Dong}}, \bibinfo {author} {\bibfnamefont {Z.}~\bibnamefont {Fan}}, \bibinfo
  {author} {\bibfnamefont {L.}~\bibnamefont {Shi}}, \bibinfo {author}
  {\bibfnamefont {A.}~\bibnamefont {Harju}},\ and\ \bibinfo {author}
  {\bibfnamefont {T.}~\bibnamefont {Ala-Nissila}},\ }\bibfield  {title}
  {\bibinfo {title} {Equivalence of the equilibrium and the nonequilibrium
  molecular dynamics methods for thermal conductivity calculations: From bulk
  to nanowire silicon},\ }\href {https://doi.org/10.1103/PhysRevB.97.094305}
  {\bibfield  {journal} {\bibinfo  {journal} {Physical Review B}\ }\textbf
  {\bibinfo {volume} {97}},\ \bibinfo {pages} {094305} (\bibinfo {year}
  {2018})}\BibitemShut {NoStop}%
\bibitem [{\citenamefont {Jahn}\ and\ \citenamefont
  {Madden}(2007)}]{ref:Jahn2007}%
  \BibitemOpen
  \bibfield  {author} {\bibinfo {author} {\bibfnamefont {S.}~\bibnamefont
  {Jahn}}\ and\ \bibinfo {author} {\bibfnamefont {P.~A.}\ \bibnamefont
  {Madden}},\ }\bibfield  {title} {\bibinfo {title} {Modeling earth materials
  from crustal to lower mantle conditions: A transferable set of interaction
  potentials for the cmas system},\ }\href
  {https://doi.org/https://doi.org/10.1016/j.pepi.2007.04.002} {\bibfield
  {journal} {\bibinfo  {journal} {Physics of the Earth and Planetary
  Interiors}\ }\textbf {\bibinfo {volume} {162}},\ \bibinfo {pages} {129}
  (\bibinfo {year} {2007})}\BibitemShut {NoStop}%
\bibitem [{\citenamefont {Slack}\ and\ \citenamefont
  {Ross}(1985)}]{ref:slack1985}%
  \BibitemOpen
  \bibfield  {author} {\bibinfo {author} {\bibfnamefont {G.~A.}\ \bibnamefont
  {Slack}}\ and\ \bibinfo {author} {\bibfnamefont {R.~G.}\ \bibnamefont
  {Ross}},\ }\bibfield  {title} {\bibinfo {title} {Thermal conductivity under
  pressure and through phase transitions in solid alkali halides. ii. theory},\
  }\href {https://doi.org/10.1088/0022-3719/18/20/021} {\bibfield  {journal}
  {\bibinfo  {journal} {J. Phys. C: Solid State Phys.}\ }\textbf {\bibinfo
  {volume} {18}},\ \bibinfo {pages} {3957} (\bibinfo {year}
  {1985})}\BibitemShut {NoStop}%
\bibitem [{\citenamefont {Manthilake}\ \emph {et~al.}(2011)\citenamefont
  {Manthilake}, \citenamefont {de~Koker}, \citenamefont {Frost},\ and\
  \citenamefont {McCammon}}]{ref:manthilake2011}%
  \BibitemOpen
  \bibfield  {author} {\bibinfo {author} {\bibfnamefont {G.~M.}\ \bibnamefont
  {Manthilake}}, \bibinfo {author} {\bibfnamefont {N.}~\bibnamefont
  {de~Koker}}, \bibinfo {author} {\bibfnamefont {D.~J.}\ \bibnamefont
  {Frost}},\ and\ \bibinfo {author} {\bibfnamefont {C.~A.}\ \bibnamefont
  {McCammon}},\ }\bibfield  {title} {\bibinfo {title} {Lattice thermal
  conductivity of lower mantle minerals and heat flux from earth's core},\
  }\href {https://doi.org/10.1073/pnas.1110594108} {\bibfield  {journal}
  {\bibinfo  {journal} {Proc Natl Acad Sci U S A}\ }\textbf {\bibinfo {volume}
  {108}},\ \bibinfo {pages} {17901} (\bibinfo {year} {2011})}\BibitemShut
  {NoStop}%
\bibitem [{\citenamefont {Imada}\ \emph {et~al.}(2014)\citenamefont {Imada},
  \citenamefont {Ohta}, \citenamefont {Yagi}, \citenamefont {Hirose},
  \citenamefont {Yoshida},\ and\ \citenamefont {Nagahara}}]{ref:imada2014}%
  \BibitemOpen
  \bibfield  {author} {\bibinfo {author} {\bibfnamefont {S.}~\bibnamefont
  {Imada}}, \bibinfo {author} {\bibfnamefont {K.}~\bibnamefont {Ohta}},
  \bibinfo {author} {\bibfnamefont {T.}~\bibnamefont {Yagi}}, \bibinfo {author}
  {\bibfnamefont {K.}~\bibnamefont {Hirose}}, \bibinfo {author} {\bibfnamefont
  {H.}~\bibnamefont {Yoshida}},\ and\ \bibinfo {author} {\bibfnamefont
  {H.}~\bibnamefont {Nagahara}},\ }\bibfield  {title} {\bibinfo {title}
  {Measurements of lattice thermal conductivity of mgo to core-mantle boundary
  pressures},\ }\href {https://doi.org/10.1002/2014gl060423} {\bibfield
  {journal} {\bibinfo  {journal} {Geophysical Research Letters}\ }\textbf
  {\bibinfo {volume} {41}},\ \bibinfo {pages} {4542} (\bibinfo {year}
  {2014})}\BibitemShut {NoStop}%
\bibitem [{\citenamefont {Roufosse}\ and\ \citenamefont
  {Klemens}(1973)}]{ref:roufosse1973}%
  \BibitemOpen
  \bibfield  {author} {\bibinfo {author} {\bibfnamefont {M.}~\bibnamefont
  {Roufosse}}\ and\ \bibinfo {author} {\bibfnamefont {P.~G.}\ \bibnamefont
  {Klemens}},\ }\bibfield  {title} {\bibinfo {title} {Thermal conductivity of
  complex dielectric crystals},\ }\href
  {https://doi.org/10.1103/PhysRevB.7.5379} {\bibfield  {journal} {\bibinfo
  {journal} {Physical Review B}\ }\textbf {\bibinfo {volume} {7}},\ \bibinfo
  {pages} {5379} (\bibinfo {year} {1973})}\BibitemShut {NoStop}%
\bibitem [{\citenamefont {Feng}\ and\ \citenamefont
  {Ruan}(2016)}]{ref:Feng2016}%
  \BibitemOpen
  \bibfield  {author} {\bibinfo {author} {\bibfnamefont {T.}~\bibnamefont
  {Feng}}\ and\ \bibinfo {author} {\bibfnamefont {X.}~\bibnamefont {Ruan}},\
  }\bibfield  {title} {\bibinfo {title} {Quantum mechanical prediction of
  four-phonon scattering rates and reduced thermal conductivity of solids},\
  }\href {https://doi.org/10.1103/PhysRevB.93.045202} {\bibfield  {journal}
  {\bibinfo  {journal} {Phys. Rev. B}\ }\textbf {\bibinfo {volume} {93}},\
  \bibinfo {pages} {045202} (\bibinfo {year} {2016})}\BibitemShut {NoStop}%
\bibitem [{\citenamefont {Wagner}\ \emph {et~al.}(2011)\citenamefont {Wagner},
  \citenamefont {Sohl}, \citenamefont {Hussmann}, \citenamefont {Grott},\ and\
  \citenamefont {Rauer}}]{ref:wagner2011}%
  \BibitemOpen
  \bibfield  {author} {\bibinfo {author} {\bibfnamefont {F.~W.}\ \bibnamefont
  {Wagner}}, \bibinfo {author} {\bibfnamefont {F.}~\bibnamefont {Sohl}},
  \bibinfo {author} {\bibfnamefont {H.}~\bibnamefont {Hussmann}}, \bibinfo
  {author} {\bibfnamefont {M.}~\bibnamefont {Grott}},\ and\ \bibinfo {author}
  {\bibfnamefont {H.}~\bibnamefont {Rauer}},\ }\bibfield  {title} {\bibinfo
  {title} {Interior structure models of solid exoplanets using material laws in
  the infinite pressure limit},\ }\href
  {https://doi.org/10.1016/j.icarus.2011.05.027} {\bibfield  {journal}
  {\bibinfo  {journal} {Icarus}\ }\textbf {\bibinfo {volume} {214}},\ \bibinfo
  {pages} {366} (\bibinfo {year} {2011})}\BibitemShut {NoStop}%
\bibitem [{\citenamefont {Dziewonski}\ and\ \citenamefont
  {Anderson}(1981)}]{ref:dziewonski1981}%
  \BibitemOpen
  \bibfield  {author} {\bibinfo {author} {\bibfnamefont {A.~M.}\ \bibnamefont
  {Dziewonski}}\ and\ \bibinfo {author} {\bibfnamefont {D.~L.}\ \bibnamefont
  {Anderson}},\ }\bibfield  {title} {\bibinfo {title} {Preliminary reference
  earth model},\ }\href {https://doi.org/10.1016/0031-9201(81)90046-7}
  {\bibfield  {journal} {\bibinfo  {journal} {Physics ofthe Earth and Planetary
  Interiors}\ }\textbf {\bibinfo {volume} {25}},\ \bibinfo {pages} {297}
  (\bibinfo {year} {1981})}\BibitemShut {NoStop}%
\bibitem [{\citenamefont {Lars~Stixrude}(2007)}]{ref:stixrude2007}%
  \BibitemOpen
  \bibfield  {author} {\bibinfo {author} {\bibfnamefont {C.~L.-B.}\
  \bibnamefont {Lars~Stixrude}},\ }\bibfield  {title} {\bibinfo {title}
  {nfluence of phase transformations on lateral heterogeneity and dynamics in
  earth's mantle},\ }\href {https://doi.org/10.1016/j.epsl.2007.08.027}
  {\bibfield  {journal} {\bibinfo  {journal} {Earth and Planetary Science
  Letters}\ }\textbf {\bibinfo {volume} {263}},\ \bibinfo {pages} {45 }
  (\bibinfo {year} {2007})}\BibitemShut {NoStop}%
\end{thebibliography}%

\end{document}